\documentclass{PoS}

\title{$\tau^-\to K^-\eta^{(\prime)}\nu_\tau$: a primer analysis}

\ShortTitle{$\tau^-\to K^-\eta^{(\prime)}\nu_\tau$: a primer analysis}

\author{
\speaker{Rafel Escribano}
\thanks{I would like to express my gratitude to the HADRON 13 Organizing Committee
for the opportunity of presenting this contribution, and for the pleasant and interesting workshop we have enjoyed.}\\
Grup de F\'{\i}sica Te\`orica (Departament de F\'{\i}sica) and Institut de F\'{\i}sica d'Altes Energies (IFAE),
Universitat Aut\`onoma de Barcelona, E-08193 Bellaterra (Barcelona), Spain\\
E-mail: \email{rescriba@ifae.es}
}

\abstract{
The decays $\tau^-\to K^-\eta\nu_\tau$ and $K^-\eta^\prime\nu_\tau$ are studied in the context of Chiral Perturbation Theory
supplemented with the explicit inclusion of resonances.
For the required vector-form factors we have explored three different possibilities according to the treatment of final-state interactions:
Breit-Wigner parametrisation, exponential resummation and dispersive representation.
In the first part of the analysis, we predict the invariant mass spectrum and the integrated branching ratio for both decays from a fit to
data on $\tau^-\to K_S\pi^-\nu_\tau$ decays.
In the second part, we take advantage of existing data on $\tau^-\to K^-\eta\nu_\tau$ from BABAR and Belle collaborations
and perform a fit to extract the pole position of the $K^\ast(1410)$ resonance which is seen to be in agreement and competitive
with the values obtained before.
From the comparison to data we conclude that the Breit-Wigner parametrisation of form factors is too na\"{\i}ve and consequently
the use of more advanced treatments such as exponential resummation or dispersive representation is essential.
}

\FullConference{
XV International Conference on Hadron Spectroscopy-Hadron 2013\\
4-8 November 2013\\
Nara, Japan}

\begin{document}
\section{Introduction}
In a paper of 2009 by the Belle Collaboration \cite{Inami:2008ar},
after briefly commenting on the different existing calculations of various $\tau$ branching ratios involving $\eta$ meson(s) in the context of chiral theories,
it is explicitly said \emph{``Further detailed studies of the physical dynamics in $\tau$ decays with $\eta$ mesons are required''}.
A proper treatment of these decays is of vital importance for the TAUOLA program, the standard Monte-Carlo generator for $\tau$ decays.
Hadronic decays of the $\tau$ (several decay modes including an $\eta$ meson represent a wide class of such decays)
are essential for studying QCD phenomena at a low-energy scale.
It is then the purpose of this contribution, which is based on the published paper in Ref.~\cite{Escribano:2013bca},
to provide a comprehensive analysis of the $\tau^-\to K^-\eta\nu_\tau$ and $K^-\eta^\prime\nu_\tau$ decays, which are still poorly studied,
emphasising the different treatments of final-state interactions.
From the point of view of theory, the first of these decays has been considered among others by Pich \cite{Pich:1987qq},
with an estimated branching ratio of $1.2\times 10^{-4}$ calculated in Chiral Perturbation Theory at lowest order, 
and Li \cite{Li:1996md}, $2.2\times 10^{-4}$ in a Vector Meson Dominance framework.
No previous analysis existed for the $K^-\eta^\prime$ decay.
On the experimental side, the invariant mass distribution and the branching ratio of the $K^-\eta$ decay has been measured recently by
Belle \cite{Inami:2008ar}, $(1.58\pm 0.05\pm 0.09)\times 10^{-4}$, 
and \textit{BABAR} \cite{delAmoSanchez:2010pc}, $(1.42\pm 0.11\pm 0.07)\times 10^{-4}$.
For the $K^-\eta^\prime$ decay only an upper bound from \textit{BABAR} \cite{Lees:2012ks} exists, $<2.4\times 10^{-6}$ at 90\% CL.

The setup of our approach to describe the $\tau^-\to K^-\eta\nu_\tau$ and $K^-\eta^\prime\nu_\tau$ decays is the following.
The different parameterisations of the required $K\eta^{(\prime)}$ vector-form factors (VFFs) depending on the treatment of final-state interactions (FSI)
are discussed below for the illustrative case of the $K\pi$ VFF.
They are calculated in the context of Resonance Chiral Theory (RChT) taking into account the effects of the $K^\ast(892)$ and $K^\ast(1410)$ vector resonances
(see Ref.~\cite{Escribano:2013bca} for a compilation of formul\ae\ and numerics).
The relative weight between their contributions is left as a free parameter.
They are found to be $f_+^{K\eta^{(\prime)}}(s)=\cos\theta_P(\sin\theta_P)f_+^{K\pi}(s)$,
where $\theta_P$ is the $\eta$-$\eta^\prime$ mixing angle in the octet-singlet basis\footnote{
The singlet contribution to the VFFs is seen to be null.
}.
The VFFs obtained in this way satisfy the chiral and high-energy constraints.
For the $K\eta^{(\prime)}$ scalar-form factors (SFFs), which also enter into the corresponding $0\to K^-\eta^{(\prime)}$ matrix elements, 
we use the well-established results of Ref.~\cite{Jamin:2001zq}
derived from a coupled-channel dispersion-relation analysis with three channels $(K\pi, K\eta, K\eta^\prime)$.
The normalisation at zero of the $K\eta^{(\prime)}$ vector- and scalar-form factors are related,
and being so, it is usual to express the matrix elements by means of their normalised versions,
$\tilde f_{+,0}^{K\eta^{(\prime)}}(s)=f_{+,0}^{K\eta^{(\prime)}}(s)/f_{+,0}^{K\eta^{(\prime)}}(0)$,
and factorise the normalisation, which is chosen to be the one associated to the VFF.
The resulting normalisation appearing in the invariant mass spectrum,
$|V_{us}|f_+^{K\eta^{(\prime)}}(0)=\cos\theta_P(\sin\theta_P)|V_{us}|f_+^{K\pi}(0)$,
will be fixed by the averaged value of $|V_{us}|f_+^{K\pi}(0)$ from different $K_{l3}$ decays \cite{Beringer:1900zz}
and the value of the mixing angle measured by KLOE \cite{Ambrosino:2006gk}.

\section{Different treatments for final-state interactions}
To illustrate the different treatments of final-state interactions (FSI) in the vector-form factors (VFFs)
let us discuss the simplest case of the $K\pi$ VFF in the isospin limit.
In Chiral Perturbation Theory (ChPT) extended to include the pseudoscalar singlet $\eta_0$ in the large-$N_c$ limit,
the $K\pi$ VFF at next-to-leading order in the chiral expansion is
\begin{equation}
\label{VFFChPT}
\left.f_+^{K\pi}(s)\right|_{\rm ChPT}=1+\frac{2}{F^2}L_9^r\, s+\frac{3}{2}\left[\tilde H_{K\pi}(s)+\cos\theta_P\tilde H_{K\eta}(s)+\sin\theta_P\tilde H_{K\eta^\prime}(s)\right]\ ,
\end{equation}
where $F$ is a common pseudoscalar decay constant in the chiral limit, $L_9^r$ is a renormalised ${\cal O}(p^4)$ low-energy constant (LEC),
and $\tilde H_{PQ}(s)$ are two-pseudoscalars loop-functions
(whose renormalisa\-tion-scale dependence cancels out with the one present in $L_9^r$)
accounting for the unitary corrections.
In the context of Resonance Chiral Theory (RChT), where nonets of different types of resonances (vectors, axial-vectors, scalars, and pseudo scalars)
are included in addition to the lowest nonet of pseudoscalar mesons, the same $K\pi$ VFF is given by
\begin{equation}
\label{VFFRChT}
\left.f_+^{K\pi}(s)\right|_{\rm RChT}=1+\frac{F_V G_V}{F^2}\frac{s}{m_{K^\ast}^2-s}\ ,
\end{equation}
where $F_V$ and $G_V$ (only one nonet of vector resonances is assumed for illustration)
are two constants identifying the couplings of the lowest nonet of vector resonances with the pseudoscalar mesons
and $m_{K^\ast}$ is the explicit vector mass for this case.
If one imposes this VFF to vanish at $s\to\infty$ at least as $1/s$ the short-distance constraint (sdc) $F_V G_V=F^2$ is obtained
and therefore the VFF is rewritten as
\begin{equation}
\label{VFFRChTsdc}
\left.f_+^{K\pi}(s)\right|_{\rm RChT+sdc}=\frac{m_{K^\ast}^2}{m_{K^\ast}^2-s}\ .
\end{equation}
Expanding now this version of the VFF around $s=0$ and comparing the result with the polynomial part of its ChPT counterpart in Eq.~(\ref{VFFChPT})
one gets $L_9^r=F^2/2 m_{K^\ast}^2$
(needless to say, the hypothesis of resonance saturation, that is, the ${\cal O}(p^4)$ LECs are saturated by the exchange of resonances, is assumed).
The VFF in Eq.~(\ref{VFFRChTsdc}), with no FSI included, is the starting point.
How does this VFF take into account the FSI (unitary corrections)?
To answer this question, three different choices can be examined, Breit-Wigner parametrisation (BW), exponential resummation (exp), and dispersive representation (DR).
Let's discuss them in turn:
\begin{itemize}
\item
\emph{BW parametrisation}:
the imaginary part of the loop-functions $H_{PQ}(s)$ is resummed to all orders in perturbation theory and identify as an energy-dependent width
while the real part is neglected.
This identification, $-\frac{3}{2}m_{K^\ast}^2\Im \left[\tilde H_{K\pi}(s)+\cdots\right]=-m_{K^\ast}\gamma_{K^\ast}(s)$, allows to write the VFF as
\begin{equation}
\label{VFFBW}
\left.f_+^{K\pi}(s)\right|_{\rm BW}=\frac{m_{K^\ast}^2}{m_{K^\ast}^2-s-i\,m_{K^\ast}\gamma_{K^\ast}(s)}\ ,
\end{equation}
where
\begin{equation}
\label{edwidth}
\gamma_{K^\ast}(s)=\gamma_{K^\ast}\frac{s}{m_{K^\ast}^2}
\frac{\sigma_{K\pi}^3(s)+\cos^2\theta_P\sigma_{K\eta}^3(s)+\sin^2\theta_P\sigma_{K\eta^\prime}^3(s)}{\sigma_{K\pi}^3(m_{K^\ast}^2)}\ ,
\end{equation}
with
$\sigma_{PQ}(s)=\frac{2q_{PQ}(s)}{\sqrt s}\Theta\left[s-(m_P+m_Q)^2\right]$, 
$q_{PQ}(s)=\frac{\sqrt s}{2}\sqrt{1-2\Sigma_{PQ}/s+\Delta_{PQ}^2/s^2}$, $(\Sigma,\Delta)_{PQ}=m_P^2\pm m_Q^2$, and
$\gamma_{K^\ast}=\gamma_{K^\ast}(m_{K^\ast}^2)$.
The value of $\gamma_{K^\ast}$ is fixed in RChT as soon as some short-distance constraints are imposed,
however, we prefer to leave $\gamma_{K^\ast}$ as a free parameter to be fitted from data.
The VFF written in this way is not analytic, in the sense that the real part of the unitary corrections is obviated,
and only takes into account the absorptive (imaginary) part of these corrections.
\item
\emph{exponential resummation} \cite{Jamin:2006tk,Jamin:2008qg}:
the real part of the loop-functions $H_{PQ}(s)$ is resummed to all orders and represented by an exponential function
while the imaginary part is kept as before.
The VFF then reads as
\begin{equation}
\label{VFFexp}
\left.f_+^{K\pi}(s)\right|_{\rm exp}=
\frac{m_{K^\ast}^2}{m_{K^\ast}^2-s-i\,m_{K^\ast}\gamma_{K^\ast}(s)}
e^{\frac{3}{2}\Re\left[\tilde H_{K\pi}(s)+\cos\theta_P\tilde H_{K\eta}(s)+\sin\theta_P\tilde H_{K\eta^\prime}(s)\right]}\ .
\end{equation}
In this case, the VFF is not yet analytic, in the sense that the real and imaginary parts of the unitary corrections are resummed in two different functions,
but both the absorptive and dispersive (real) part of the corrections are considered.
This exponential resummation is inspired by the fact that the form factor (FF) should satisfy a dispersion relation (DR).
The DR relates the real and imaginary parts of the FF.
When the imaginary part of the FF is expressed in terms of the real part and the associated phase,
the DR becomes an integral equation of the real part whose solution is the Omnès exponential, an integral of the FF phase,
multiplied by a polynomial which takes into account possible subtraction constants 
(these constants are required to make the FF well behaved at high energies or to diminish the contribution of this high-energy part when it is not known)
to be fixed at low energies.
If, in addition, elastic unitarity is applied, the phase of the FF must be equal to the phase of the final state elastic scattering.
When this equality is implemented and the lowest order result in ChPT is used for the scattering phase,
the outcome of the Omnès integral is the result in Eq.~(\ref{VFFexp}) but in the elastic case, that is, the $K\pi$ channel alone.
Then, the imaginary part of the exponential is identified as the energy-dependent width of the resonance and only the real part remains as its argument
(see the details of this approach in the seminal work of Ref.~\cite{Guerrero:1997ku}).
If inelasticities were not important, the exponential resummation should be a good representation of the FF.
However, in our case, where we want to consider the $K\eta$ VFF as well, two channels have to be taken into account at least, $K\pi$ and $K\eta$,
and elastic unitarity cannot be employed.
Therefore, the representation in Eq.~(\ref{VFFexp}) is neither totally unitary, although many unitary corrections are included,
nor completely analytic (for the reasons explained above), but still could be a good representation of the FF.
\item
\emph{dispersive representation} \cite{Boito:2008fq,Boito:2010me}:
the real and imaginary parts of the loop-functions $H_{PQ}(s)$ are resummed and both accommodated in the resonance propagator.
The VFF is then given by
\begin{equation}
\label{VFFDRinput}
\left.f_+^{K\pi}(s)\right|_{\rm DR}^{\rm input}=
\frac{m_{K^\ast}^2}{m_{K^\ast}^2-s-\frac{3}{2}m_{K^\ast}^2\Re\left[\tilde H_{K\pi}(s)+\cdots\right]-i\,m_{K^\ast}\gamma_{K^\ast}(s)}\ .
\end{equation}
In this case, the VFF is analytic and the unitary corrections are included in a perturbative way as before.
Compared to the previous case, the two expressions of the VFF are equal at next-to-leading order but differ at next-to-next-to-leading order.
From the VFF in Eq.~(\ref{VFFDRinput}), one extracts its phase through $\phi_{K\pi}^{\rm input}(s)\equiv\arctan\left[\Im f_+^{K\pi}(s)/\Re f_+^{K\pi}(s)\right]$
and this is inserted into a three-times-subtracted dispersive representation of the FF
\begin{equation}
\label{VFFDRoutput}
\left.f_+^{K\pi}(s)\right|_{\rm DR}^{\rm output}=f_+^{K\pi}(0)
\exp\left[\alpha_1\frac{s}{m_\pi^2}+\frac{1}{2}\alpha_2\frac{s^2}{m_\pi^4}+
\frac{s^3}{\pi}\int_{s_{K\pi}}^{s_{\rm cut}}ds^\prime\frac{\phi_{K\pi}^{\rm input}(s^\prime)}{(s^\prime)^3(s^\prime-s-i\epsilon)}\right]\ ,
\end{equation}
where the constants $\alpha_{(1,2)}$ are related to the linear and quadratic slope parameters $\lambda^{(\prime,\prime\prime)}_+$ of the FF\footnote{
The Taylor expansion of the FF around $s=0$ is given by 
$f_+^{K\pi}(s)=f_+^{K\pi}(0)\left(1+\lambda^\prime_+\frac{s}{m_\pi^2}+\frac{1}{2}\lambda^{\prime\prime}_+\frac{s^2}{m_\pi^4}+
\frac{1}{6}\lambda^{\prime\prime\prime}_+\frac{s^3}{m_\pi^6}+\cdots\right)$.
}
through $\lambda^\prime_+=\alpha_1$ and $\lambda^{\prime\prime}_+=\alpha_2+\alpha_1^2$, respectively, $s_{K\pi}=(m_K+m_\pi)^2$, and
$s_{\rm cut}$ is a cut-off whose value is fixed from the requirement that the fitted parameters do not change within errors when compared to the case
$s_{\rm cut}\to\infty$.
This particular representation of the VFF is chosen to reduce the high-energy contribution to the phase integral
(obviously, $\phi_{K\pi}^{\rm input}(s)$ cannot be the correct phase from threshold to infinity).
For the case at hand, the study of $\tau\to(K\pi,K\eta)\nu_\tau$ decays, the value $s_{\rm cut}=4$ GeV$^2$ is seen to satisfy this criterion\footnote{
In practise, the value of the cut-off is varied in order to estimate the associated systematic error.}.
As in the previous case, the VFF in Eq.~(\ref{VFFDRoutput}) is not fully unitary, however, its phase is seen to agree quite reasonably with the
elastic $K\pi$ scattering phase up to the first inelastic threshold located just above 1 GeV, thus pointing out that most of the unitary corrections are taken into account.
One disadvantage of the description of the VFF in terms of a three-times-subtracted DR is that its proper behaviour at high energy is not guaranteed
because the subtraction constants will be fixed from a fit to experimental data.
However, the values obtained for these constants permit the FF to follow the expected behaviour.
\end{itemize}

\section{Predictions for $\tau^-\to K^-\eta\nu_\tau$ and $K^-\eta^\prime\nu_\tau$ decays}
\begin{figure}[pt]
\centerline{
\includegraphics[width=0.5\textwidth]{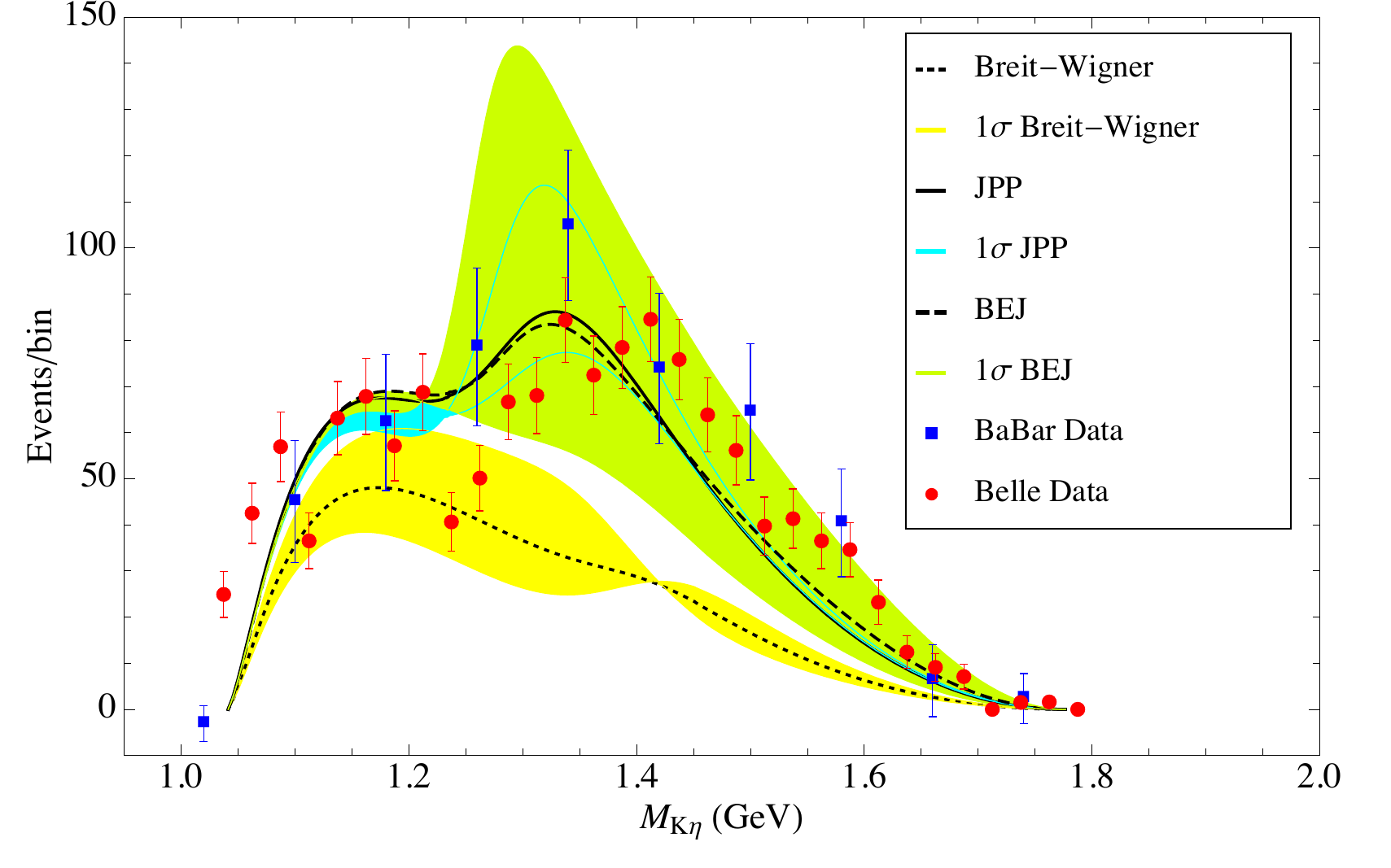}
\includegraphics[width=0.5\textwidth]{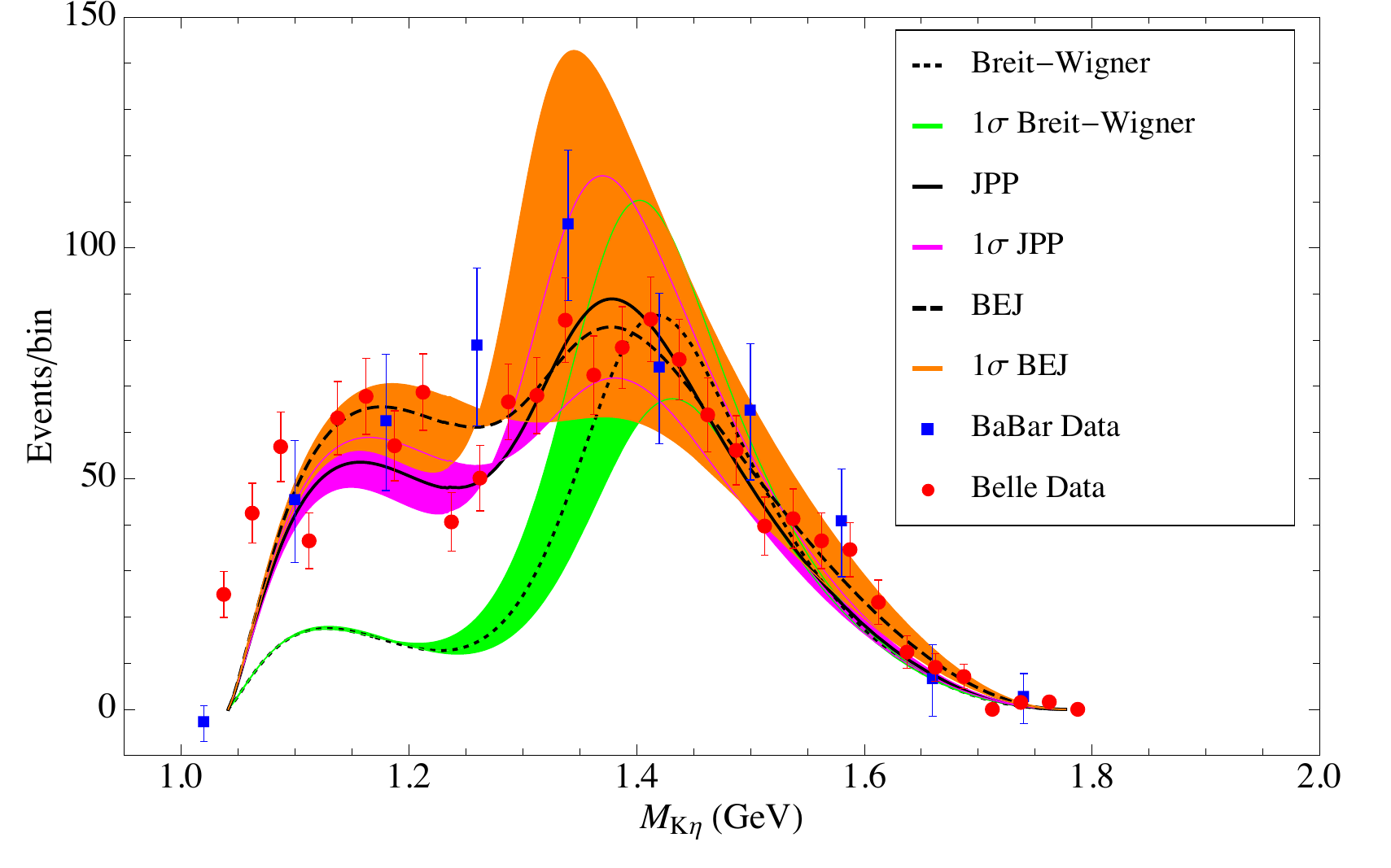}
}
\caption{
\textit{left}\textit{(right)}:
Belle (red) and \textit{BABAR} (blue) experimental data for the $\tau^-\to K^-\eta\nu_\tau$ decays as compared to the predictions(best fits) obtained
from the BW (dotted), JPP (solid) and BEJ (dashed) representations of the $K\eta$ VFF.
The corresponding $1\sigma$-error bands are presented in yellow(light green), light blue(pink) and light green(orange), respectively.
\label{figKetaspectrum}
}
\end{figure}

The invariant mass spectrum of the $\tau^-\to K^-\eta\nu_\tau$ and $K^-\eta^\prime\nu_\tau$ decays
is predicted, in the cases of the exponential resummation and dispersive representation, from the results obtained 
in the analysis of $\tau^-\to K_S\pi^-\nu_\tau$ experimental data released by Belle \cite{Epifanov:2007rf}
using the same parameterisation in each case, that is, 
from Ref.~\cite{Jamin:2008qg} for the exponential resummation (JPP) and Ref.~\cite{Boito:2010me} for the dispersive representation (BEJ).
For the BW parameterisation, however, we use the masses and widths of the $K^\ast(892)$ and $K^\ast(1410)$ vector resonances
published by the PDG \cite{Beringer:1900zz} and for the parameter weighting their relative contributions we used the value given by RChT.
The different predictions for the $K\eta$ invariant mass spectrum and their corresponding $1\sigma$-error bands are shown in Figure \ref{figKetaspectrum} \textit{(left)}
and compared with existing data from Belle \cite{Inami:2008ar} and \textit{BABAR}\footnote{
The \textit{BABAR} data are normalised to Belle data.} 
\cite{delAmoSanchez:2010pc}.
While the BW parameterisation of the $K\eta$ VFF fails in reproducing the measured spectrum,
the exponential resummation and the dispersive representation agree reasonably well with data.
Moreover, the integrated branching ratio in both cases are in accord with the PDG value, $(1.52\pm 0.08)\times 10^{-4}$, within errors.
For the $\tau^-\to K^-\eta^\prime\nu_\tau$ decays, we obtain in all cases values of the branching ratio around $1\times 10^{-6}$,
well below the experimental upper bound.

\section{Fit to $\tau^-\to K^-\eta\nu_\tau$ data}
How do the experimental data on the invariant mass spectrum of $\tau^-\to K^-\eta\nu_\tau$ decays affect the former predictions?
To answer this question, we use these Belle and \textit{BABAR} sets of data to fix the parameters of the different descriptions.
To simplify, we only leave as free parameters the mass and width of the $K^\ast(1410)$ together with the relative weight of the two vector resonances
($\tau^-\to K^-\eta\nu_\tau$ data is not sensitive to the mass and width of the $K^\ast(892)$ resonance).
The best fit for each description is presented in Figure \ref{figKetaspectrum} \textit{(right)}.
The fitted values for the different parameters and the pole position of the $K^\ast(1410)$ resonance obtained in each case
can be found in Ref.~\cite{Escribano:2013bca}.
Again, the two more elaborated representations of the $K\eta$ VFF agree much better with data than the simplest BW parameterisation. 
Moreover, the results obtained for the $K^\ast(1410)$ mass and width in the present analysis,
which are competitive with the values from $\tau^-\to K_S\pi^-\nu_\tau$ data,
suggest the possibility of performing a joint fit to experimental $K_S\pi^-$  and $K^-\eta$ invariant mass distributions
in order to further improve the pole position of the $K^\ast(1410)$ resonance \cite{EGSJR}.

\end{document}